\documentclass[journal=jpccck,manuscript=article]{achemso}

\usepackage[version=4]{mhchem}
\usepackage{graphicx}
\usepackage{amsmath}
\usepackage{soul}
\setkeys{acs}{doi = true}
\usepackage{hyperref}
\DeclareUnicodeCharacter{2009}{\,}
\usepackage{xcolor}

\author{Jalen Macatangay}
\affiliation{School of Materials Engineering, Purdue University, West Lafayette, Indiana, 47907, USA}
\alsoaffiliation{Theoretical Division, Los Alamos National Laboratory, Los Alamos, New Mexico, 87545, USA}
\author{Brenden W. Hamilton}
\affiliation{Theoretical Division, Los Alamos National Laboratory, Los Alamos, New Mexico, 87545, USA}
\author{Alejandro Strachan}
\email{strachan@purdue.edu}
\affiliation{School of Materials Engineering, Purdue University, West Lafayette, Indiana, 47907, USA}

\title{Polymer-Mediated Hotspot Evolution in Shock-Induced Pore Collapse in 1,3,5,7-Tetranitro-1,3,5,7-tetrazocane (HMX)-based Explosives}

\begin{document}

\begin{abstract}
    The initiation of detonation in high explosives (HEs) under shock loading is strongly influenced by the collapse of porosity and the subsequent formation of hotspots. As these materials are widely integrated with polymer binders to form polymer-bonded explosives, the collapse of pores at explosive-binder interfaces has received significant interest. Recent molecular dynamics (MD) simulations revealed that the shock-driven expansion of soft polymers (e.g. polystyrene) into a void produces greater pressure-volume (PV) work during recompression and higher local temperatures compared to pore collapse in neat HEs, ultimately expediting the onset of chemical reactions. Yet how pore collapse, interfacial phenomena, and binder chemistry jointly affect initiation remains insufficiently understood. We provide deeper insights into the role of polymer in hotspot evolution through reactive MD simulations, specifically characterizing its chemical behavior during the shock-induced collapse of voids at sizes approaching experimentally relevant length scales. We consider the HE 1,3,5,7-Tetranitro-1,3,5,7-tetrazocane (HMX) in a crystal-gap-crystal stack with polystyrene along the upstream surface. In the hydrodynamic collapse regime, hotspots nucleate at the material interfaces, with void length-dependence in PV work, shock temperature, and the onset of self-sustaining chemistry. Driven by shock-induced deformation and HMX decomposition, the binder undergoes backbone chain scission followed by extensive aromatic-backbone cross-bonding, akin to a graphite to diamond transformation. These reactions act as heat sinks, suppressing deflagration and ultimately inhibiting hotspot growth and coalescence between crystals. The simulations further link the polymer’s chemical response to the shock-to-detonation transition in polymer-bonded explosives.
\end{abstract}

\newpage
\section*{Introduction}

The shock-induced initiation leading to detonation in high explosives (HEs) involves a complex interplay of thermal, mechanical, and chemical processes that occur over a wide range of length and time scales \cite{Handley2018UnderstandingScales,Hamilton2021ChemistryConditions}. Initial chemical reactions behind the shock front are governed by interactions between the shockwave and microstructural inhomogeneities, concentrating energy in regions referred to as hotspots \cite{Mader1965InitiationDiscontinuities}. These localized areas of elevated temperature can ultimately facilitate deflagration, depending on the size, temperature, and degree of molecular deformation \cite{Tarver1996CriticalExplosives,Hamilton2021APhysics}. While various mechanisms are known to generate hotspots \cite{Chaudhri1974TheExplosives,Dienes2006ImpactMechanics,Bowden1949InitiationGrit,Winter1975TheExplosives,Field1992HotExplosives,Dattelbaum2010InfluenceNitromethane,Kroonblawd2026AnisotropicCrystal}, the collapse of porosity is known to dominate HE shock initiation. This was demonstrated by the desensitization of explosives following weak shocks that eliminated porosity without triggering reactions \cite{Campbell1985ShockB-3}.

The extremely short time (ps) and length (nm$-\mu$m) scales of shock loading and hotspot formation have underscored the importance of molecular dynamics (MD) simulations. Combined with reactive force fields (i.e. ReaxFF), MD has shed light on the underlying atomic-scale mechanisms that connect pore collapse mechanics and chemical decomposition to hotspot criticality \cite{Strachan2003ShockRDX,Zhou2014ShockSimulation,Wood2015UltrafastNanoscale,Shan2016ExtendedExplosive,Islam2020RoleDynamics,Hamilton2025ThermalHMX}. The dynamics of pore collapse are governed by a wide range of coupled physical processes mediated by the loading conditions and void geometry. Under weak shocks, the collapse of cylindrical/spherical pores is viscoplastic, characterized by pressure-driven incompressible plastic flow and viscous heating \cite{Carroll1972StaticMaterials}. Above a critical shock strength, hydrodynamic collapse dominates, during which the upstream material focuses into a fluid-like jet that can outpace the incident shock \cite{Mader1965InitiationDiscontinuities,Eason2015Molecular-RDX,Wood2018MultiscaleMaterial,Nguyen2026Three-dimensionalDynamics}. Hotspot temperatures then become strongly influenced by pressure-volume (PV) work arising from recompression against the downstream pore wall. Within the hydrodynamic regime, PV work can be amplified with larger voids \cite{Holian2002AtomisticInitiation,Kroonblawd2025ScaleExplosives} and high-curvature regions, which enhance shock focusing and jetting \cite{Li2020HotspotTemperature,Li2022Shock-inducedTemperature}. Coupled with this, other factors such as crystallographic structure \cite{Sakano2018RoleRDX,Li2021Shock-inducedStudy}, anisotropy \cite{Grilli2018TheMaterials,Zhao2021StronglyTATB,Hamilton2022TheOrientation}, shearing \cite{Islam2020RoleDynamics,Izvekov2022Bottom-up-RDX,Izvekov2022MicroscopicTransition,Hamilton2023IntergranularWork,Hamilton2024HighFormation,Hamilton2025ThermalRDX,Wilkening2025Steady-stateCrystal}, and mechanochemistry \cite{Kroonblawd2020HighBands,Hamilton2021APhysics,Hamilton2022ExtemporaneousEnergy,Hamilton2023Many-bodyChemistry,Michalchuk2024OnSolids}, further contribute to chemical initiation.

During the collapse of voids with complex geometries, shock focusing, jetting, and intricate material interactions result in complex hotspot shapes that hinder a comprehensive understanding of microstructural effects on hotspot dynamics \cite{Hamilton2023UsingSolids}. Hence, planar (quasi-1D) voids often serve as a reductionist approach to directly map PV work to energy localization while closely resembling intergranular configurations in polycrystalline materials \cite{Hamilton2023IntergranularWork}. This allows for a direct mapping of effects such as crystal orientation, material composition, and thermodynamic conditions onto PV evolution. Within the hydrodynamic collapse regime \cite{Mader1965InitiationDiscontinuities}, upstream material accelerates to velocities up to twice the particle velocity of the shock ($U_p$) \cite{Forbes2012ShockMatter}. Holian et al. showed that the critical shock strength and void size associated with the vaporization of ejecta material can be analytically determined and is directly related to the cohesive energy of the material \cite{Holian2002AtomisticInitiation}. The resulting PV work produces heating well above the initial shock temperature, with the expected maximum heating ($\Delta T_{max}$) directly proportional to the shock ($U_s$) and particle velocity: $k_B\Delta T_{max}=mU_sU_p/d$ \cite{Holian2002AtomisticInitiation}.

Despite significant research efforts over the past decade, many MD simulations still neglect the local effects of polymer binders, which are relevant in most explosive formulations \cite{Li2022SystematicExplosives}. Understanding the role of binder on hotspot criticality in polymer-bonded explosives (PBXs) remains a significant modeling challenge. The microstructures of PBXs are structurally complex, characterized by large binder-HE interfacial surface areas and heterogeneous porosity \cite{Mang2011SmallAngleIntensity}. These interfaces give rise to impedance mismatches which produce reflective shockwaves that can alter hotspot temperatures depending on the interfacial configuration \cite{Menikoff2011HotReflections,An2014InhibitionInterface,Hamilton2023InfluenceLocalization}. The binder formulation also affects interfacial structure and properties \cite{Yeager2012NanoindentationFailure}, particularly to deformation processes such as delamination, which can introduce additional void formation, microcracking, and hotspots \cite{Yeager2014Formulation-derivedMaterials,Manner2017InModeling,You2015UltrasonicSolids}. 

Under shock loading, the response of polymers is influenced by viscoelasticity, strain-rate, and chemical reactions, even for inert compositions \cite{Clements2012ALoading}. Distinctive relaxation mechanisms such as dynamic glass transitions, virtual melting, and chain restructuring also occur over an extensive range of timescales (ps - $\mu$s) \cite{Veysset2017MolecularElastomers,Lecoutre2021HugoniostatMelts,Macatangay2022DeviatoricPolymers}. At high shock pressures (P $>$ 20$-$30 GPa) \cite{Carter1995HugoniotPolymers}, chemical decomposition becomes the dominant response mechanism, mediated by backbone degradation and dissociation reactions \cite{Mattsson2010First-principlesPolymers,Morris1984TheGPa,Bowman2021SupersonicSimulations,Hamilton2023RapidSpallation}. Specifically, recent simulations on shocked phenolics showed that following dehydrogenation, compression was facilitated by the combination of carbon fragments into nonaromatic structures \cite{Moore2025Shock-inducedPolymer}. With porosity, these chemical activation thresholds have been found to decrease substantially \cite{Dattelbaum2014ReactiveFoams}. Interestingly, reactive MD simulations on planar pore collapse in RDX-polystyrene systems unexpectedly revealed the acceleration of HE reactions following the shock-driven recompression of the inert polymer. This resulted from the greater expansion of the more compliant polymer, which generated increased PV work, and subsequently, higher temperature hotspots \cite{Macatangay2024InfluenceDynamics}.

While prior studies have provided insights into hotspot formation mechanisms associated with the collapse of the binder, significant research gaps remain that limit the fidelity of mesoscale PBX models. Most notably, the influence of shock-induced polymer chemistry on hotspot growth, HE decomposition, and deflagration propagation has received little attention at the molecular scale. Under shock loading, binder-HE interfaces are known to drive localized heating \cite{Dandekar2019TheVelocities} and even generate intense hotspots \cite{You2015UltrasonicSolids}. However, their interplay with porosity to govern hotspot formation and subsequent chemistry remains largely unexplored. Furthermore, the effect of void size at the binder-HE interfaces on critical hotspot formation and growth is not fully understood, despite its crucial role in initiation probabilities and run to detonation distances \cite{Dattelbaum2010InfluenceNitromethane,Herring2010EffectsExplosive}. Ultra-small angle neutron scattering experiments on PBXs observed void sizes ranging from 10$-$1000 nm \cite{Peterson2005QuantitativeGradient,LeePerry2018RelatingSensitivity}. Specifically, in pristine PBX 9501, while most pores sizes are concentrated around 40 nm, additional sizes of $\sim$200-400 nm exist at the binder-HE interfaces, with the total void volume dominated by voids in the $\sim$500 nm range \cite{Mang2002An9501}. While MD has been capable of simulating 100 nm-scale void collapse, such studies only consider pure HE systems \cite{Li2022Shock-inducedTemperature,Kober2024AtomisticHMX,Kroonblawd2025ScaleExplosives}. Hence, we conducted reactive MD simulations of shock-induced planar pore collapse, which directly enable quantification of the polymer's role in pore collapse, hotspot formation, and reactive chemistry. To improve length-scale comparisons with experiments, we focus on sub-micron crystals of 1,3,5,7-tetranitro-1,3,5,7-tetrazocane (HMX), the primary component of PBX 9501, separated by gaps up to 200 nm with a polystyrene (PS) film situated along the upstream pore surface. While the collapse of larger voids increases the PV work done by the polymer, leading to faster shock-to-deflagration transition and hotspot growth, they also induce substantial PS decomposition that obstructs crystal-to-crystal deflagration propagation. The simulations elucidate the coupled mechanistic and chemical processes that influence energy localization, reaction initiation, and deflagration growth in inert-binder PBX microstructures.

\section*{Methods}

Reactive molecular dynamics (MD) simulations were conducted using the LAMMPS software package \cite{Thompson2022LAMMPSScales}, with atomic bonding and interactions described by the 2018 parameterization of the ReaxFF force field (ReaxFF-2018) \cite{Wood2018MultiscaleMaterial}. This corrects long-range London dispersion interactions through the inclusion of a low-gradient term \cite{Liu2011ReaxFF-Materials}, resulting in more accurate densities and detonation properties \cite{Islam2019ReactiveNitromethane,Hamilton2023High-pressureMaterials}. All simulations were performed using a 0.1 fs time step, with partial atomic charges calculated with a charge equilibration (QEq) tolerance of $1 \times 10^{-6}$ \cite{Rappe1991ChargeSimulations}. To create the planar gap system, we first produce a supercell through a 5x4x4 replication of the $\beta$-HMX unit cell \cite{Cady1963The-HMX}. After initial equilibration under isothermal-isobaric (NPT) conditions at 300 K and 1 atm, the cell was then replicated 400 times in the shock direction (z). Nonperiodic boundary conditions were applied in the shock direction, while periodic conditions were retained in the lateral directions. Two HMX slabs with lengths of 1000 nm and 200 nm were then formed by removing $\sim$70 nm of bulk material in unit cell proportions of molecules. The resulting breathing modes due to the creation of free surfaces were then attenuated through 200 ps of thermalization under isothermal-isochoric (NVT) conditions at 300 K. We divided the first and last 50 ps into five sets, each set consisting of 2.5 ps of a Langevin thermostat \cite{DUNWEG1991BrownianNumbers} and 7.5 ps of a Nose-Hoover thermostat \cite{Nose1984AMethods}, both with 100 fs damping parameters.

\begin{figure}
    \centering
    \includegraphics[width=1\textwidth]{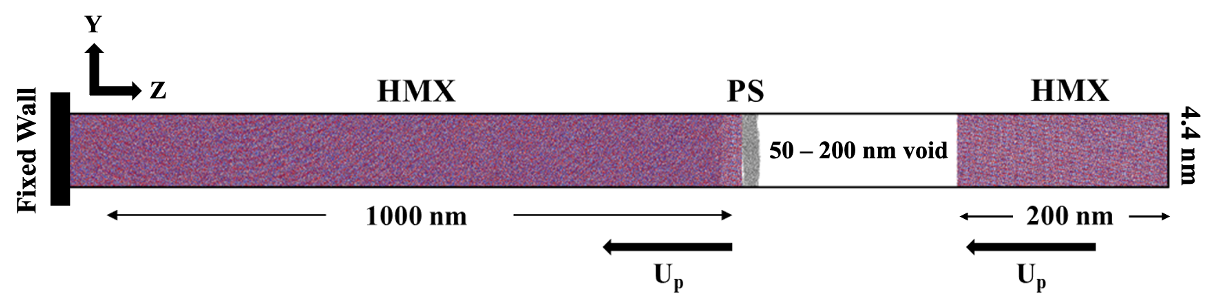}
    \caption{Shock-induced planar pore collapse simulation setups with a particle velocity ($U_p$) of 2 km/s and gap sizes between 50 and 200 nm.}
    \label{fig:model}
\end{figure}

The polystyrene (PS) structure was generated using the Polymer Modeler tool available on nanoHUB \cite{Haley2015AtomisticPolymerModeler} via the continuous configuration bias Monte Carlo method \cite{Siepmann1992ConfigurationalChains}. Polymer chains consisted of 150 monomers, corresponding to a molecular weight of $\sim$15620 g/mol that exceeds the PS entanglement length ($\sim$13300 g/mol) \cite{Fetters1994ConnectionProperties}. A bulk system with 27 chains was first equilibrated at 800 K and 1 atm using the Dreiding force field \cite{Mayo1990DREIDING:Simulations}, until a constant density was obtained ($\sim$0.74 g/cm\textsuperscript{3}). The lateral dimensions were additionally deformed during the relaxation to conform to the cross-sectional dimensions of the HMX crystal. Subsequently, the system was cooled to 300 K, decreasing the temperature by 10 K every 200 ps. Free surfaces were then introduced by removing chains with molecules outside a $\sim$20 nm region, followed by energy minimization and NVT relaxation using ReaxFF-2018. The resulting density is around 1.08 g/cm\textsuperscript{3}, which is in reasonable agreement with experiments (1.05 g/cm\textsuperscript{3}) \cite{Carter1995HugoniotPolymers,Alzate-Vargas2018UncertaintiesDynamics}. We placed the PS slab along the upstream surface of the gap and performed additional NVT to equilibrate the resulting interface. Systems with gap lengths of 50, 100, 150, and 200 nm were then created through outward translation of the downstream HMX crystal. Figure~\ref{fig:model} shows the final HMX-PS planar gap system, which contains 2.7 million HMX and 36,030 PS atoms. We additionally constructed HMX-only systems of identical gap lengths to further evaluate how the polymer affects shock-induced heating and reactivity across these larger void collapse cases.

Shocks simulations were conducted under adiabatic (NVE) conditions through the reverse ballistic approach \cite{Holian1979MolecularConditions}. Specifically, the left surface (z = 0) behaves as a reflective boundary, and all atoms are assigned an extra z-velocity equal to the particle velocity ($U_p$) directed leftwards. As a result, the fully supported shockwave travels along the +z direction at velocity $U_s-U_p$, with the shocked material nearly stationary. Through Galilean invariance, this is identical to shocks generated by a moving piston impacting material initially at rest. We perform shocks with $U_p$ = 2 km/s, which is well in the hydrodynamic collapse regime and above the critical shock strength \cite{Macatangay2024InfluenceDynamics}. Spatial profiles of temperature, pressure, and density were calculated at every 0.1 ps as an average of atoms within fixed 1 nm-thick Eulerian bins along the shock direction. We also track PV work and chemistry for regions of material closest to the impact surface through a Lagrangian bin framework, dividing into bins of approximately 2500 atoms, each consisting of only HMX or only PS.

Chemical species and bonds were determined through bond order calculations every 0.1 ps using a cutoff of 0.5. For deeper insights into individual reactions, particularly within the polymer, we additionally employ a coordinate geometry analysis (CGA), tracking the types of atoms bonded to a particular central atom \cite{Kober2022DevelopingTATB,Sakano2020UnsupervisedRDX}. As an example, a PS monomer would be described as follows, where the element outside the brackets denotes the central atom: 8 H[C], 5 C[HCC], 1 C[CCC], 1 C[HCCC], 1 C[HHCC]. This approach is advantageous in reducing the complexity of tracking large numbers of molecules and has been utilized to develop reduced-order chemical kinetics models for HEs \cite{Sakano2020UnsupervisedRDX,Lafourcade2023MultiscaleLearning} and assess first step reaction pathways \cite{Hamilton2021PredictedLLM-105,Hamilton2022ExtemporaneousEnergy}.

\section*{Results and Discussion}

\subsection*{Overall Shock Response and Hotspot Evolution}

\begin{figure}
    \centering
    \includegraphics[width=1\textwidth]{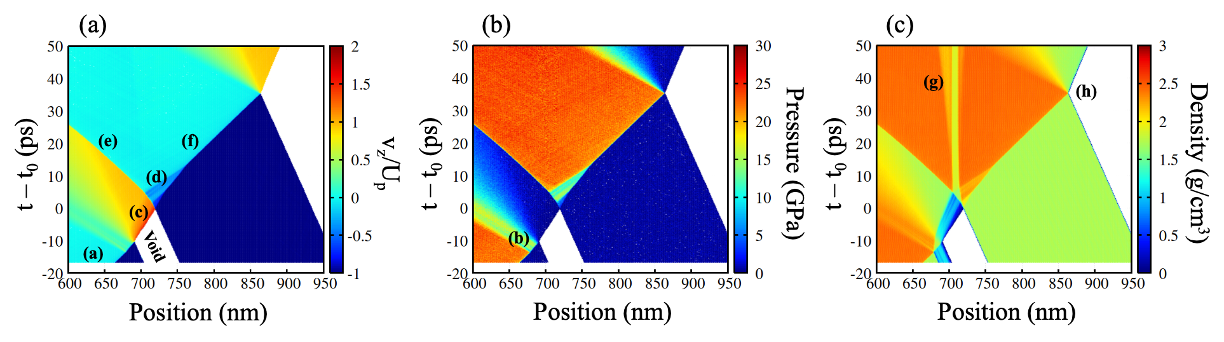}
    \caption{\textit{x–t} diagrams relative to the pore collapse time ($t_0$) for the 50 nm void system of (a) atomic velocity in the shock direction ($v_z$) normalized by $U_p$ = 2 km/s, (b) pressure, and (c) density.}
    \label{fig:xt_50}
\end{figure}

The shock-induced collapse of porosity in energetic composites results in a heterogeneous response characterized by the production and interaction of multiple waves propagating in different directions, governed by the constituent materials' physical properties. We illustrate this progression over time and position (referred to as \textit{x-t} diagrams) relative to the time at which the leading portion of the polymer impacts the downstream crystal ($t_0$). Figure~\ref{fig:xt_50} shows \textit{x-t} diagrams of atomic velocity ($v_z$), pressure, and density for the $U_p$ = 2 km/s, 50 nm gap simulation. We describe the primary stages of pore collapse as follows. First, the incident shockwave generated from the fixed wall propagates through the upstream HMX crystal, raising temperatures and pressures to approximately 760 K and 23 GPa (a). As it transitions into the lower-impedance polymer, a reflective wave forms that partially decompresses the shocked HMX to around 15 GPa (b). Meanwhile, the transmitted wave reaches the free surface, propelling the upstream material into the void as an expanding rarefaction wave, reaching velocities close to 1.5$U_p$ (c). The polymer then collides with the downstream crystal at the impact surface (d), generating compressive shockwaves in both directions. These are further intensified through interactions with the reflected (e) and transmitted (f) waves arising from the collapse of the higher-impedance HMX, leading to a maximum compression pressure of $\sim$25 GPa. From the density \textit{x-t} diagram, the PS film can be discerned, achieving densities as low as 0.1 g/cm\textsuperscript{3} immediately before impact. The lighter shaded regions adjacent to the polymer (g) correspond to significant exothermic, volume-expanding reactions, signifying the onset of deflagration. Lastly, at roughly $t_0$ + 35 ps, the shockwave reaches the free surface at the far end of the system (h).

\begin{figure}
    \centering
    \includegraphics[width=1\textwidth]{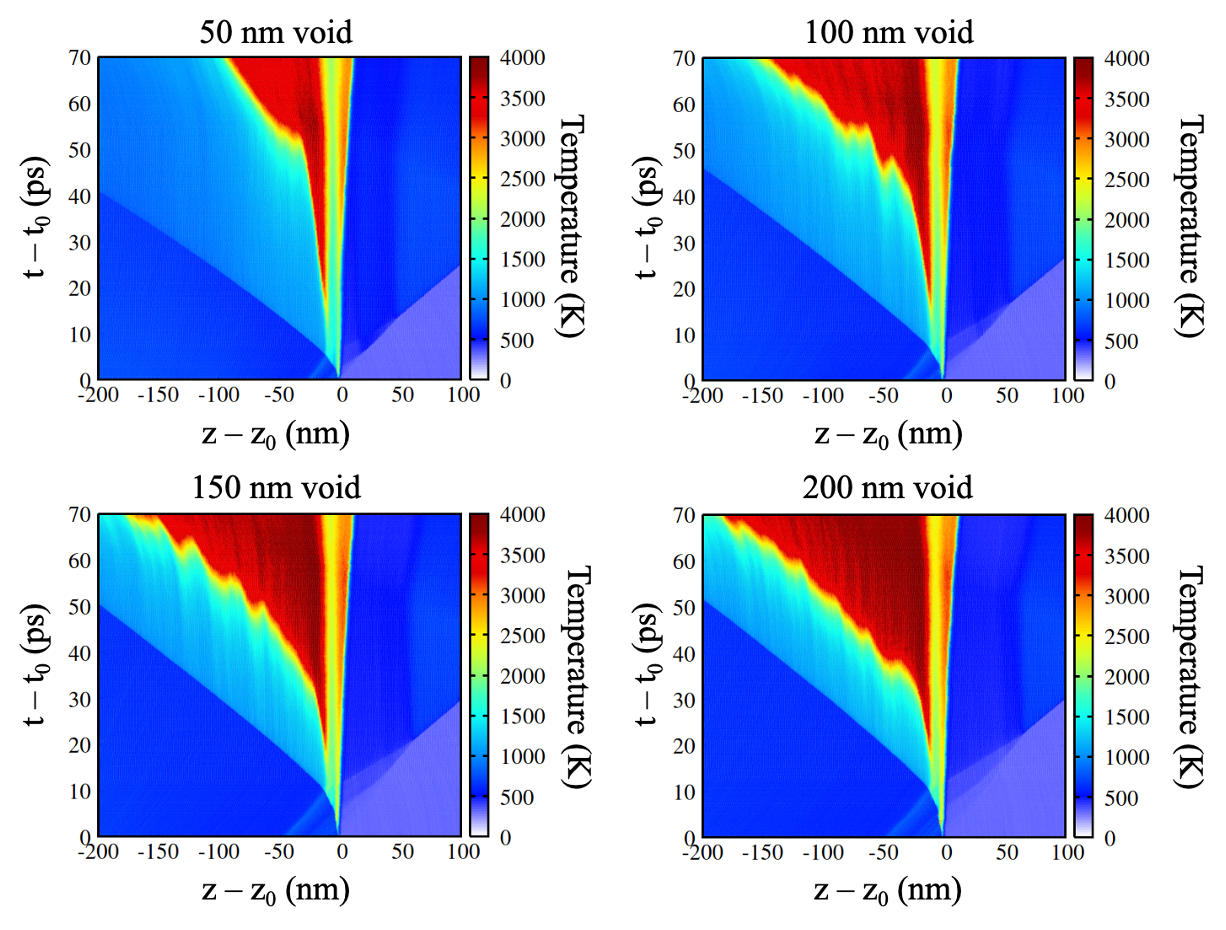}
    \caption{Temperature \textit{x–t} diagrams relative to the pore collapse time ($t_0$) and downstream interface ($z_0$) for gap sizes of 50$-$200 nm at $U_p$ = 2 km/s.}
    \label{fig:xt_temperature}
\end{figure}

The influence of PS on the hydrodynamic collapse of larger, experimentally comparable voids \cite{Mang2002An9501} is first assessed through analysis of hotspot temperatures near the collapse region. Temperature \textit{x-t} diagrams relative to $t_0$ and the downstream interface ($z_0$) are displayed in Figure~\ref{fig:xt_temperature} for all void sizes up to 200 nm. We additionally include the corresponding velocity, pressure, and density diagrams in Section SM-1 of the Supporting Information. As expected, all cases exhibit rapid onset of exothermic reactions, indicated by temperatures exceeding 3500 K and reduced densities within the hotspot region (see Figure S3 in the Supporting Information). Following pore collapse, hotspots appear to form in two distinct regions. The first originates at $z_0$, coinciding with the collision between the polymer jet against the downstream HMX crystal. The second then appears at the opposite interface following recompression. For the former, we observe that in the 50 nm void case, the impact surface immediately elevates to a temperature close to 2000 K, which then gradually rises to 3000 K in our simulation timescales. For larger void sizes, the continued rarefaction of the polymer drives higher-energy impacts. We find that for the 200 nm void case, the collision promptly elevates the impact region temperature to $\sim$2200 K, which then quickly rises to 2500 K as the polymer further impinges on the downstream wall. At $t_0$ + 40 ps, the temperature of this region reaches 3000 K and is roughly constant for the remainder of the simulation. From the density \textit{x-t} diagrams (Figure S3), small molecule products appear in all cases, most clearly in the 150 and 200 nm simulations, yet remain confined to a narrow, slow-growing region adjacent to the downstream interface. While it is not surprising that the PS film inhibits chemical initiation in the bulk downstream crystal \cite{Hamilton2023InfluenceLocalization,Macatangay2024InfluenceDynamics}, we see that sufficiently intense polymer impact can still drive localized HE reactions, albeit insufficient to propagate a deflagration wave.

Following impact, the rebounding shockwave then recompresses the polymer and subsequently the previously shocked HMX that has filled the void. Formation of the second hotspot is driven by PV work, with shock temperatures of around 1500 K and 1100 K for bulk PS and HMX, respectively. Interestingly, a localized region adjacent to the upstream interface quickly matches and surpasses the temperature of the polymer, exceeding 3000 K within 20 ps after collapse. The reshock temperatures marginally increase with larger voids, but the timescales for significant exothermic chemistry relative to $t_0$ remain roughly consistent (15$-$20 ps after $t_0$). At this point, a deflagration front promptly spreads into the upstream crystal. For the 50 nm void case, there is a stagnation time of around 50 ps before the deflagration accelerates, which diminishes in time with increasing void size. The deflagration profiles of the 150 and 200 nm cases are closely similar, indicating that the heating due to PV work is approaching its maximum, asymptotic value discussed by Holian et al. \cite{Holian2002AtomisticInitiation}. We further quantify the expansion and PV work done by the polymer, as well as its influence on local heating in the next section.

\begin{figure}
    \centering
    \includegraphics[width=1\textwidth]{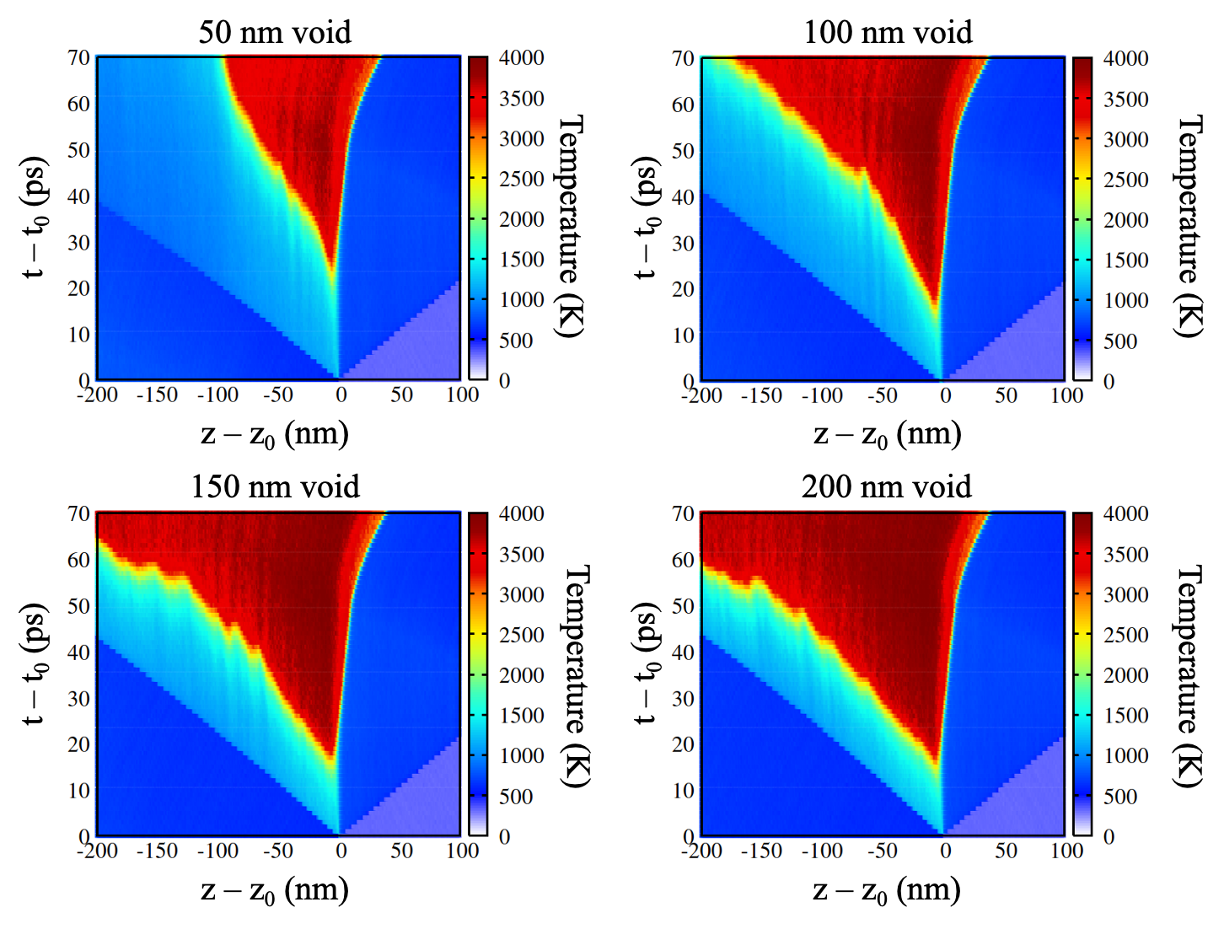}
    \caption{Temperature \textit{x-t} diagrams relative to the pore collapse time ($t_0$) and downstream interface ($z_0$) for gap sizes of 50$-$200 nm on a pure HMX (no polymer) system at $U_p$ = 2 km/s.}
    \label{fig:xt_temperature_noPS}
\end{figure}

Lastly, we compare these simulations to identical cases without polymer (HMX-only), with temperature \textit{x-t} diagrams shown in Figure~\ref{fig:xt_temperature_noPS}. Leading expansion of the denser HMX leads to comparable shock temperatures across void size (1280$-$1370 K), but lower relative to those of the polymer in the HMX-PS cases. The core of the hotspot then forms approximately at $z_0$. Similar to previous work \cite{Macatangay2024InfluenceDynamics}, the 50 nm void case takes slightly longer to initiate exothermic reactions compared to its respective HMX-PS simulation. However, for larger voids, we find that the onset of steady-state deflagration roughly aligns in time with those in the polymer containing cases. This indicates that for these cases (150 and 200 nm), the polymer has minimal effect on the formation of the upstream hotspot. The most pronounced differences, however, arise during deflagration. In the HMX-only cases (Figure~\ref{fig:xt_temperature_noPS}), the resulting deflagration front propagates freely and more rapidly into the downstream crystal. In contrast, as shown in Figure~\ref{fig:xt_temperature}, despite significant HMX decomposition, the PS film suppresses deflagration propagation into the downstream region, effectively behaving as a thermal barrier. This is directly relevant to PBX sensitivity. Although the binder does not prevent hotspot formation or exothermic chemical reactions, it can substantially delay reaction zone development by retarding deflagration propagation between crystals. The thermal and chemical mechanisms underlying this behavior are investigated in the following sections.

\subsection*{Pressure-Volume Work and Interfacial Heating}

\begin{figure}
    \centering
    \includegraphics[width=1\textwidth]{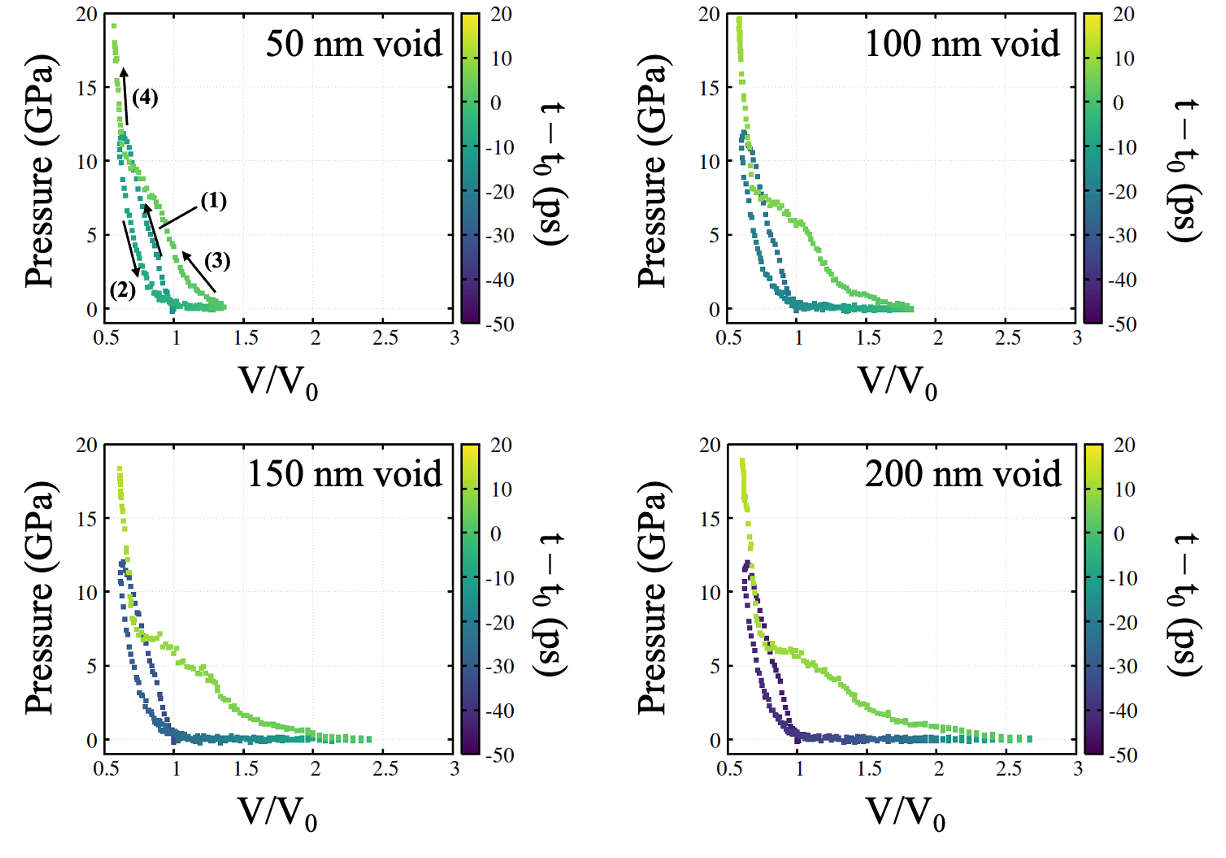}
    \caption{Pressure-volume (PV) curves of PS with gap sizes from 50 to 200 nm at $U_p$ = 2 km/s. The color scheme represents the time-evolution in PV space where lighter colors correlate with later simulation times.}
    \label{fig:work}
\end{figure}

To gain deeper insights into the effect of the polymer's PV work done on hotspot temperatures, we turn to the progression of pressure-volume states during the collapse sequence. This is shown in Figure~\ref{fig:work} as an average over all PS atoms, analogous to treating the polymer as a single, Lagrangian bin. As presented in the 50 nm case, the PV trajectory exhibits multiple stages: (1) compression due to passage of the incident shock, (2) expansion into the void, (3) recompression against the downstream wall, and (4) further compaction from impact of the higher-impedance HMX. We display the recompression P-V paths up to maximum compression. Additionally, corresponding temperature-volume (TV) diagrams are shown Section SM-2 of the Supporting Information to illustrate the contribution of PV work to temperature evolution throughout pore collapse. Assuming vapor-ejecta recompression, the temperature rise ($\Delta T=T-T_{(1)}$, where $T_{(1)}$ is the principal shock temperature) can also be analytically determined from the work done, represented in Figure~\ref{fig:work} by the area between the release isentrope (2) and the reshock Rayleigh lines (3) and (4).

Beginning with the 50 nm void simulation, polymer expansion is considerably impeded, relative to the other cases, reaching 1.5 times its initial volume. We find that its recompression occurs quickly, with a sharp reshock Rayleigh line (3) that comes close to the incident shock pressure (1). Upon increasing the void size to 100 nm, the polymer nearly reaches twice its initial volume due to expansion. In the larger void sizes (150 and 200 nm), we observe pronounced disentanglement and stretching of the leading chains, driving expansion to more than 2.5 times its initial volume. This causes recompression to unfold over longer durations, with large volume reduction corresponding to the packing of the individual chains, followed by compaction of the less rarefied "bulk" polymer region. The corresponding reshock Rayleigh lines exhibit shallower slopes, highlighting the significant densification of the polymer because of enhanced rarefaction. Lastly, regardless of void size, the polymer undergoes a maximum compression of 0.6 times its initial volume, as a result of recompression of the upstream HMX (4). As displayed in Figure S4 of the Supporting Information, higher $\Delta T$ values ($\sim$920 K and $\sim$1400 K for the 50 and 200 nm cases, respectively) indicate that the total PV work done steadily increases with void size. To compare with the case of idealized hotspot formation, described via isentropic expansion and vapor-ejecta recompression, we turn to the theoretical model formulated by Holian et al. \cite{Holian2002AtomisticInitiation}. For $U_p$ = 2 km/s, the maximum temperature rise $\Delta T_{max}$ of PS is 3020 K. From our simulations, the average temperature rise due to collapse of a 50 and 200 nm planar gap are around 30\% and 45\% of $\Delta T_{max}$. Quite interestingly, this demonstrates that increases in the void size significantly enhance expansion and thermal localization of the PS film relative to the theoretical maximum hotspot temperature. With higher aspect ratio voids, such as elongated cracks, we expect shock focusing effects to further enhance PV work, producing temperature rises closer to $\Delta T_{max}$ \cite{Li2022Shock-inducedTemperature}.

\begin{figure}
    \centering
    \includegraphics[width=1\textwidth]{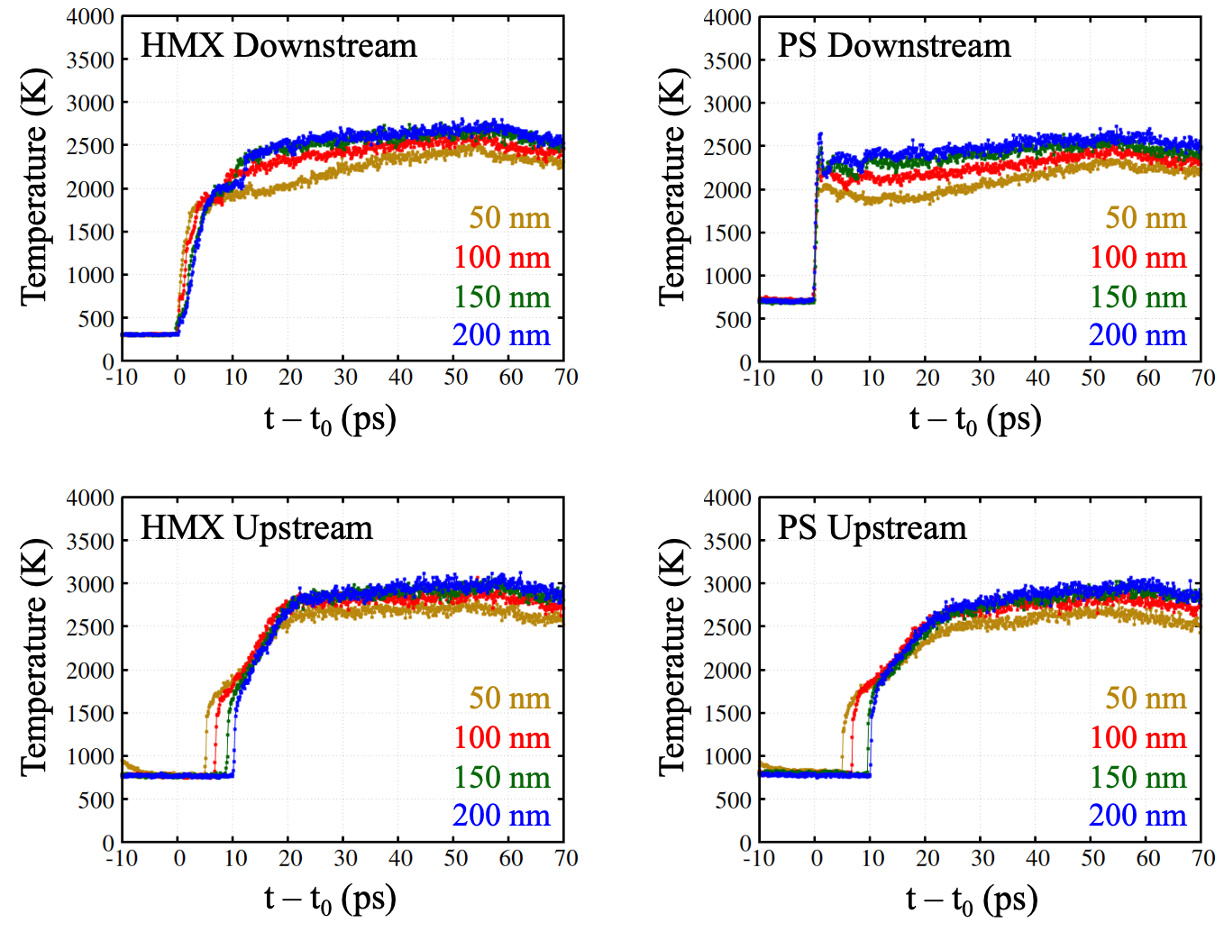}
    \caption{Temperature-time profiles of HMX and PS at the downstream (top) and upstream (bottom) interfaces for gap sizes from 50 to 200 nm. Each curve corresponds to the Lagrangian bin closest to the respective interface.}
    \label{fig:temperature_time}
\end{figure}

While pore collapse is widely recognized as a primary mechanism for chemical initiation, energy localization at interfaces has also been observed to influence local heating and reactivity \cite{Dandekar2019TheVelocities,Holbrook2025ModelingComposites,Kroonblawd2026AnisotropicCrystal}. We resolve the interfacial thermal behavior by concentrating on four distinct regions: HMX downstream, HMX upstream, PS downstream, and PS upstream. Specifically, each region corresponds to the Lagrangian bin immediately adjacent to its respective interface (downstream or upstream). Figure~\ref{fig:temperature_time} shows the temperature evolution relative to $t_0$ of these interfacial regions. We also display this for a series of bins for the 200 nm void case in Section SM-3 of the Supporting Information to identify spatial differences caused by the spreading rarefaction wave.

At the downstream interface, heating is primarily driven by collision of the polymer against the second HMX crystal. As highlighted in the "PS downstream" figure, continuous acceleration of the polymer jet notably increases impact temperatures as the void size lengthens. For the 200 nm case, the high-velocity collision generates a transient temperature rise close to 2700 K, before quickly equilibrating to $\sim$2300 K. We note that this may reflect center of mass subtraction artifacts in the calculation of bin temperatures and should thus be treated as an upper bound on the local temperature. Consistent with earlier results, the close agreement between the 150 and 200 nm profiles highlights the maximum heating from impact under these loading conditions. In contrast, for the 50 nm case, reduced jet expansion results in lower peak temperatures ($\sim$2000 K). Following maximum compression, the temperatures gradually increase by a few hundred Kelvin, possibly indicating heat conduction from the neighboring HMX. As for "HMX Downstream", larger voids produce progressively slower shock heating during collapse, but ultimately reach higher maximum compression temperatures due to greater PS expansion. In the 200 nm case, this value slightly exceeds 2500 K, which falls well short of temperatures characteristic of significant product formation and sustained deflagration ($\sim$4000 K) \cite{Wood2015UltrafastNanoscale}.

As previously discussed, the initial shock temperatures at the upstream interface are influenced by PV work. The 150 and 200 nm cases again display similar upstream temperature profiles for both HMX and PS, with the 100 nm case now more closely aligned. During recompression (first 10 ps after $t_0$), the temperature rises between "HMX Upstream" and "PS Upstream" are nearly identical. Product forming reactions then further elevate the upstream HMX temperatures close to 3000 K. Interestingly, as seen in Figure S5 in the Supporting Information, the final temperatures rise even higher for HMX regions farther from the interface, reaching values consistent with the HMX-only case (Figure~\ref{fig:xt_temperature}) and similar reactive pore collapse studies \cite{Wood2015UltrafastNanoscale,Macatangay2024InfluenceDynamics}. On the other side, the upstream PS temperatures steadily increase, albeit slower than the upstream HMX, reaching a temperature $\sim$2800 K that exceeds those at the downstream interface. For the polymer regions away from the interfaces (see Figure S5), the opposite trend is observed relative to HMX, in which the temperature remain below the interfacial values ($\sim$2500 K). These results suggest that the polymer, through endothermic chemical reactions, thermal energy absorption, or both, acts as a heat sink at the interfaces. As shown in Figure~\ref{fig:xt_temperature}, while the thermal buffering of the polymer has little influence on the upstream hotspot, particularly for larger voids, we find that this instead substantially suppresses significant exothermic chemistry in the downstream HMX crystal.

\subsection*{Interfacial and Polymer Chemistry}

\begin{figure}
    \centering
    \includegraphics[width=1\textwidth]{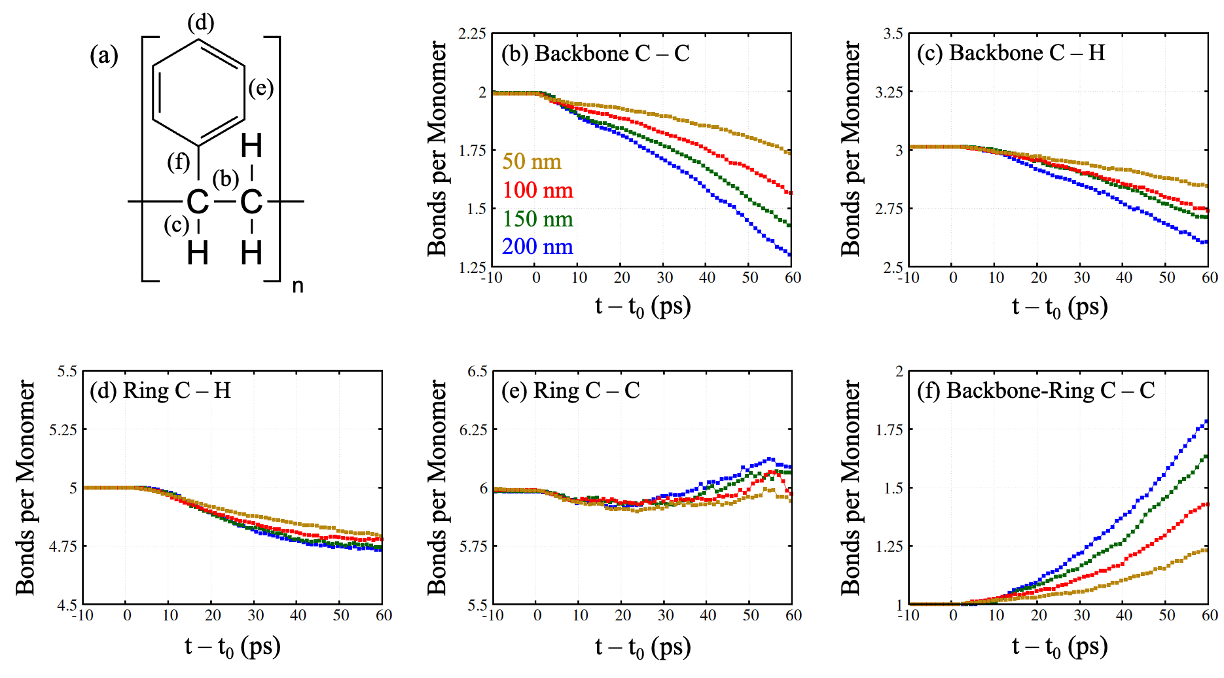}
    \caption{(a) Chemical structure of styrene with bonds labeled by type. (b)$-$(f) Time-profiles of the number of bonds per monomer for each bond type: (b) carbon-carbon backbone, (c) carbon-hydrogen backbone, (d) carbon-hydrogen ring, (e) carbon-carbon ring, (f) carbon-carbon backbone-ring.}
    \label{fig:polymer_bonds}
\end{figure}

To better characterize the transient and thermodynamic behavior at the interfaces compared to the bulk, we investigate the chemical activity of the polymer during pore collapse and its overall role in HE reactivity. We first focus on the chemical response of PS during shock-induced pore collapse as a function of void size through the evolution of all polymer bond types. This is displayed in Figure~\ref{fig:polymer_bonds} relative to $t_0$ and expressed as the number of bonds per monomer (out of 2250 total monomers). For reference, we label all bond types present in a styrene monomer in Figure~\ref{fig:polymer_bonds}(a) as follows: (b) C$-$C backbone, (c) C$-$H backbone, (d) C$-$C ring, (e) C$-$H ring, and (f) C$-$C backbone-ring. Focusing first on the backbone bonds (Figure~\ref{fig:polymer_bonds}(b) and Figure~\ref{fig:polymer_bonds}(c)), we observe considerable carbon backbone scission, accompanied by dehydrogenation to a lesser extent. With larger voids, bond scission increases across all backbone types. Specifically in the 200 nm case, each monomer lost an average of approximately 0.15 C$-$C backbone bonds (7.5\% of initial) in the 10 ps after $t_0$, associated with the expansion and recompression of the polymer. Following this, elevated polymer temperatures drive further decomposition. By $t_0$ + 60 ps, each monomer breaks on average $\sim$0.75 C$-$C backbone bonds (37.5\% of initial) and $\sim$0.40 C$-$H backbone bonds (13\% of initial).

Now turning to the benzene rings, which contain stronger bonds due to their sp$^2$ hybridized carbon, we find that the void size plays a smaller role on degradation. As shown in Figure~\ref{fig:polymer_bonds}(d), approximately 0.25 C$-$H ring bonds per monomer (5\% of initial) are lost by $t_0$ + 60 ps regardless of void size. The number of C$-$C ring bonds (Figure~\ref{fig:polymer_bonds}(e)) slightly decreases for all cases directly after impact, which then gradually reform to an extent that varies with void size. This is indicative of ring-opening reactions induced during recompression, leading to aliphatic crosslinking with neighboring ring structures, consistent with observations in strongly shocked benzene \cite{Martinez2019ParallelBenzene}. Remarkably, as seen in Figure~\ref{fig:polymer_bonds}(f), the number of C$-$C bonds between an initially backbone carbon and an initially ring carbon substantially increase. This is driven by backbone fragmentation, which similarly becomes more significant as the void size lengthens. In combination with the reformation of C$-$C ring bonds, this suggests an increase in the hybridization of the ring carbon atoms, in line with reactions seen in polymers at high shock pressures and analogous to a graphite-to-diamond phase transition \cite{Carter1995HugoniotPolymers,Moore2025Shock-inducedPolymer}. Section SM-4 of the Supporting Information further confirms this carbon hybridization transformation by tracking the populations of sp$^2$ and sp$^3$ carbon bonding environments over time and position for the 200 nm simulation.

\begin{figure}
    \centering
    \includegraphics[width=1\textwidth]{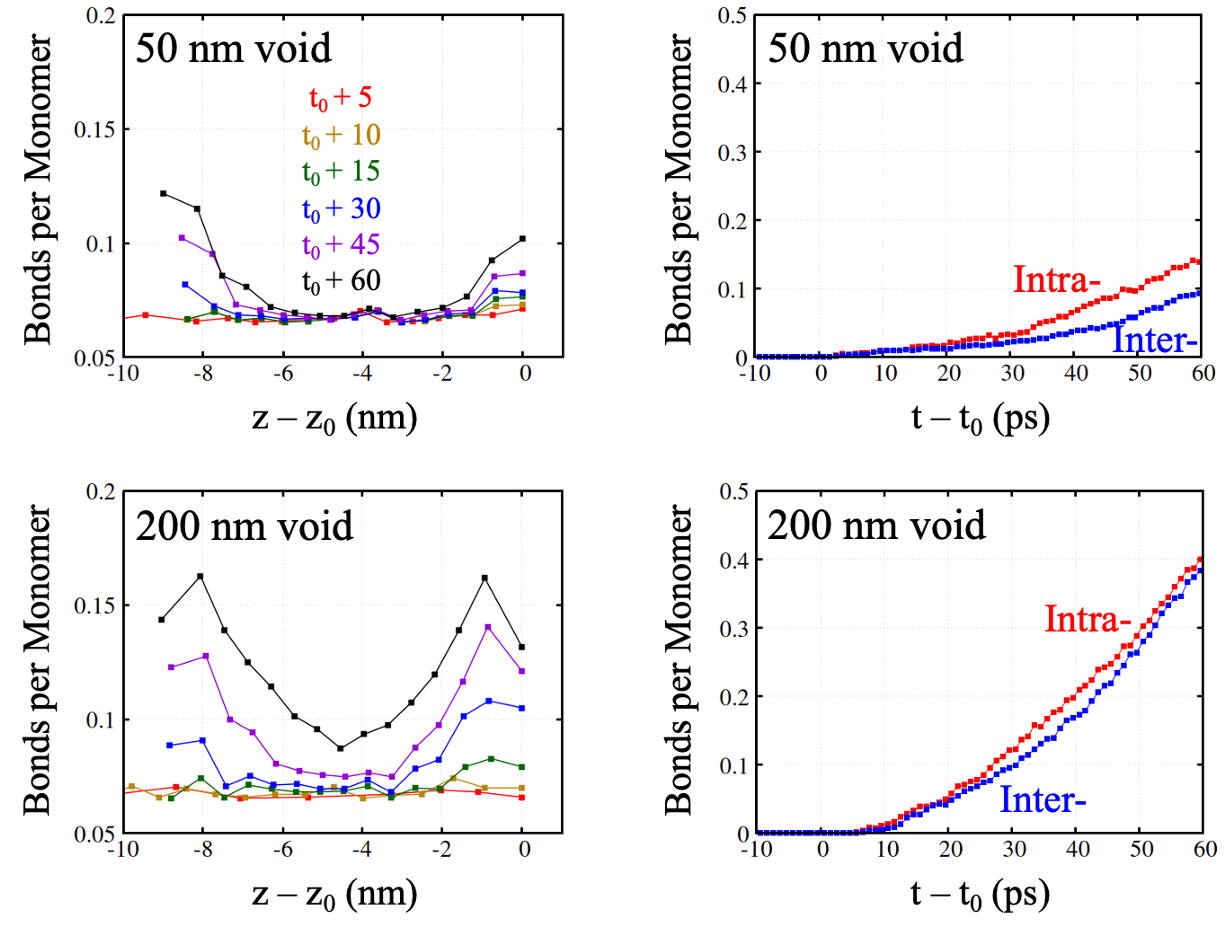}
    \caption{(left) Backbone-ring carbon bonds per monomer, displayed over position relative to the downstream interface ($z-z_{0}$) at various times following pore collapse for the 50 and 200 nm cases. (right) Intra-chain and inter-chain contributions to the newly formed C$-$C bonds, shown over time relative to $t_0$ for the same cases.}
    \label{fig:ccbr_bonds}
\end{figure}

From Figure~\ref{fig:polymer_bonds}, it is evident that the pore collapse driven chemical evolution of PS is dominated by backbone scission followed by combination with benzene rings. We now focus specifically on the backbone-ring C$-$C bonding to further elucidate the pore collapse driven carbon hybridization changes. First, in the left panels of Figure~\ref{fig:ccbr_bonds}, the bonding profile is resolved over position with respect to the downstream interface ($z-z_{0}$) at various times after pore collapse for the 50 and 200 nm void simulations. Similar figures for the 100 and 150 nm cases are provided in Section SM-5 of the Supporting Information. Consistent with earlier discussion, backbone-ring carbon bond formation is most significant at the interfaces. For both cases, while impact against the downstream HMX crystal induces backbone chain scission, minimal C$-$C backbone-ring bonds are formed at early times, as observed in the rightmost points of the $t_0$ + 5 and $t_0$ + 10 ps curves. These instead begin to significantly form as hotspots develop in the adjacent HMX. In the 50 nm case, this occurs roughly at $t_0$ + 30 ps, which then continue to form as the downstream interface further heats. At the upstream interface, bond formation is nearly double that at the downstream interface at $t_0$ + 60 ps, further emphasizing the dominant role of PV work in HMX initiation. Negligible bonds form in the central polymer region despite temperatures eventually matching that of the downstream region, attributed to a lesser degree of mechanochemical driven backbone bond rupture and reduced heating from HMX chemistry.

As for the 200 nm case, stronger collisions with the downstream surface and greater PV work lead to considerably more bond formation at both interfaces. At the downstream interface, greater backbone scission driven by impact drives rapid bond formation beginning at $t_0$ + 15 ps. While the upstream interface recompresses at a slightly later time, its bond profile quickly matches that of the downstream interface by $t_0$ + 60 ps. Relative to the 50 nm case, the points nearest the interface show a slight decrease in bonds, most likely reflecting the formation of smaller hydrocarbon molecules as a result of high interfacial temperatures or cross-bonding with HMX atoms. These products are more typical reactions for strongly shocked polymers with minimal side-chain complexity, such as polyethylene \cite{Morris1988ShockPolyethylene}. Additionally, beginning at $t_0$ + 30 ps, the C$-$C backbone-ring bonds form throughout the central polymer region. Through comparison with the 50 nm case, we see that the higher temperatures and stress localization from the collapse of larger pores play a profound role in the chemical transformation of the polymer.

To more precisely characterize the chemical mechanisms governing carbon bond formation between the backbones and rings, we revisit the time-profiles, but now separated based on the parent backbone chain of the bonded atoms. This is shown for the 50 and 200 nm case in the right panels of Figure~\ref{fig:ccbr_bonds}, where "intra-" and "inter-" denote bonds formed within the same chain and between different chains, respectively. Across all void sizes, both mechanisms contribute to bond formation. With the 50 nm case, intra-chain bonding is marginally higher, mainly associated with the lower degrees of expansion insufficient to produce significant chain intermixing. This trend is consistent, but with greater bond fractions as the void size reaches 150 nm (see Figure S7). In contrast, we find that the collapse of the 200 nm void drives relatively more cross-chain bonding, closely aligning with the intra-chain profile. The enhanced intermolecular restructuring in many polymers, especially with pedant groups, is linked to high shock pressures \cite{Carter1995HugoniotPolymers}. Yet, our simulations reveal that for shock-induced pore collapse, additional mechanisms, notably chain extension during expansion and mechanochemically induced backbone degradation \cite{Hamilton2023RapidSpallation}, further facilitate chain rearrangement for reactions.

Overall, the interface-dominant backbone-ring C$-$C clustering, which are typically volume-reducing and slightly endothermic, explains the reduced temperatures at the interfaces relative to the bulk HMX, as discussed in Figure S5 of the Supporting Information, highlighting the chemically-driven thermal dissipation behavior of the shocked polymer. This provide a mechanistic basis for the retardation of deflagration and prevention of interactions between the hotspots at the interfaces \cite{Scandolo1995Pressure-InducedDiamond,Dattelbaum2019Shock-DrivenFoams}. This can dramatically affect the deflagration-to-detonation transition, which is ultimately characterized by the coalescence of multiple hotspots \cite{Handley2018UnderstandingScales}. Although this may be unsurprising given the chemically inert nature of PS, our simulations reveal competing mechanisms under strong shocks: while increased PV work from the polymer accelerates hotspot initiation, its subsequent chemical decomposition retards hotspot growth. Both effects become more pronounced with increasing void size. 

\begin{figure}
    \centering
    \includegraphics[width=1\textwidth]{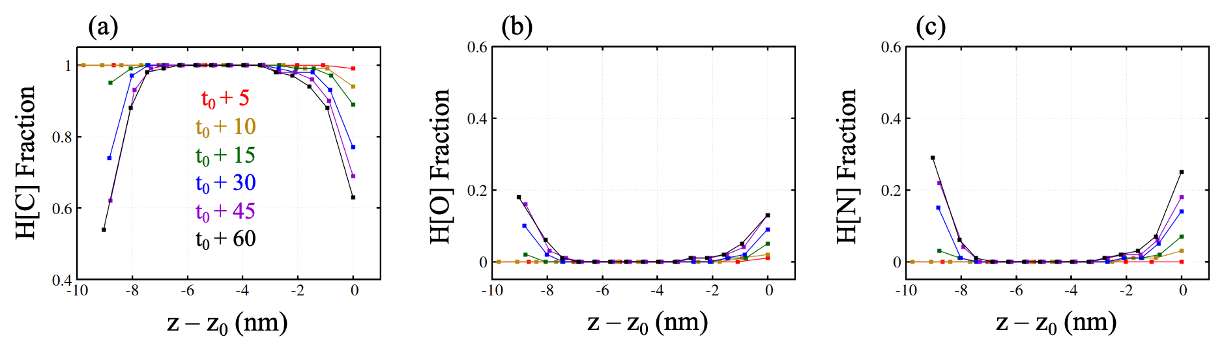}
    \caption{(a) H[C], (b) H[O], and (c) H[N] bonding environments, normalized by the initial number of hydrogen atoms per bin, as a function of position relative to the downstream interface ($z-z_{0}$) at various times following pore collapse.}
    \label{fig:H_env}
\end{figure}

Lastly, we turn our attention to the chemical activity of the PS hydrogen atoms, which have been shown to play a key role in shock-induced polymer decomposition \cite{Moore2025Shock-inducedPolymer}. Hydrogen bonding environments ((a) H[C], (b) H[O], and (c) H[N]), obtained using the CGA approach, are displayed in Figure~\ref{fig:H_env} as a function of position relative to the downstream interface and as a fraction of the initial amount of hydrogen atoms per bin. We exclude the evolution of H[H] environments, as their abundance was negligible during the simulations. As anticipated, dehydrogenation reactions are largely at the interfaces, coinciding with hotspot initiation in the adjacent HMX. Based on Figure~\ref{fig:polymer_bonds}(c) and (d), hydrogen dissociates from both the backbone, and to a lesser extent, the aromatic ring. Following, the hydrogen atoms form with oxygen and nitrogen, most likely as part of small-molecule detonation products (\ce{H2O} and \ce{NH3}). Interestingly, hydrogen more frequently bonds with nitrogen than with oxygen, possibly reflecting delayed polymer dehydrogenation at the interfaces relative to HMX oxidation. This is consistent with HMX decomposition, in which \ce{H2O} generally forms before the nitrogen-containing intermediates convert to \ce{N2} \cite{Manaa2002DecompositionSimulation,Zhang2009CarbonSimulations}. Within the bulk PS region, we see constant H[C] environments following pore collapse, unsurprisingly indicating negligible hydrogen chemistry.

\section*{Conclusions}

The polymer binder significantly influences the microstructure, mechanical response, and overall performance of PBXs, necessitating accurate representation in predictive models. Using reactive molecular dynamics simulations, we elucidate the chemical and mechanical response of an inert polymer (PS) adjacent to a planar gap between HMX crystals during shock-induced pore collapse. Notably, our simulations capture binder decompositions reactions in the collapse of voids of lengths up to 200 nm, consistent with the sub-micron pore distributions in PBX 9501. We find that, under hydrodynamic collapse ($U_p$ = 2 km/s), larger voids enable greater polymer expansion, leading to highly intense collisions that localize significant energy at the downstream binder-HE interface. Following, recompression of the expanded material generates substantial PV work, which heightens with void size and promptly triggers deflagration into the upstream material. Collapse of the larger voids induces extensive binder degradation, characterized by backbone fragmentation and the combination of hydrocarbon fragments into non-aromatic sp$^3$ carbon networks. These reactions are typically driven by high shock pressures, yet our simulations show that pore collapse and elevated interfacial temperatures also serve as primary mechanisms. We also observe that the endothermic, volume-collapsing nature of PS decomposition suppresses deflagration into the downstream HMX, ultimately preventing hotspot growth between crystals. Overall, these results demonstrate that even an inert binder substantially influences PBX shock initiation through coupled mechanical and chemical processes that can both accelerate and delay stages of the deflagration-to-detonation transition. Further reactive simulations of binders with varying chemical structure and properties under shock loading are necessary to provide accurate atomistic insights into binder selection, as well as improve predictability of MD informed models of PBX shock-to-detonation transition.

\begin{acknowledgement}
    Funding was provided by the Advanced Simulation and Computing Physics and Engineering Models project (ASC-PEM).
    Partial funding was provided by the Army Research Office and was accomplished under Cooperative Agreement No. W911NF-22-2-0170. The views and conclusions contained in this document are those of the authors and should not be interpreted as representing the official policies, either expressed or implied, of the Army Research Office or the U.S. Government. The U.S. Government is authorized to reproduce and distribute reprints for Government purposes notwithstanding any copyright notation herein.
    This research used resources provided by the Los Alamos National Laboratory (LANL) Institutional Computing Program.
    This work was supported by the U.S. Department of Energy (DOE) through LANL, which is operated by Triad National Security, LLC, for the National Nuclear Security Administration of the U.S. Department of Energy (Contract No. 89233218CNA000001). 
    Approved for unlimited release: LA-UR-26-29243.
\end{acknowledgement}

\begin{suppinfo}
    Additional \textit{x-t} diagrams, temperature-volume profiles during recompression, Lagrangian temperatures profiles for bins nearest the interfaces, evolution of the carbon hybridization for the 200 nm case, and backbone-ring C$-$C figures for the intermediate void size simulations
\end{suppinfo}

\bibliography{references}

\begin{tocentry}
    \includegraphics[width=\linewidth,height=1.75in,keepaspectratio]{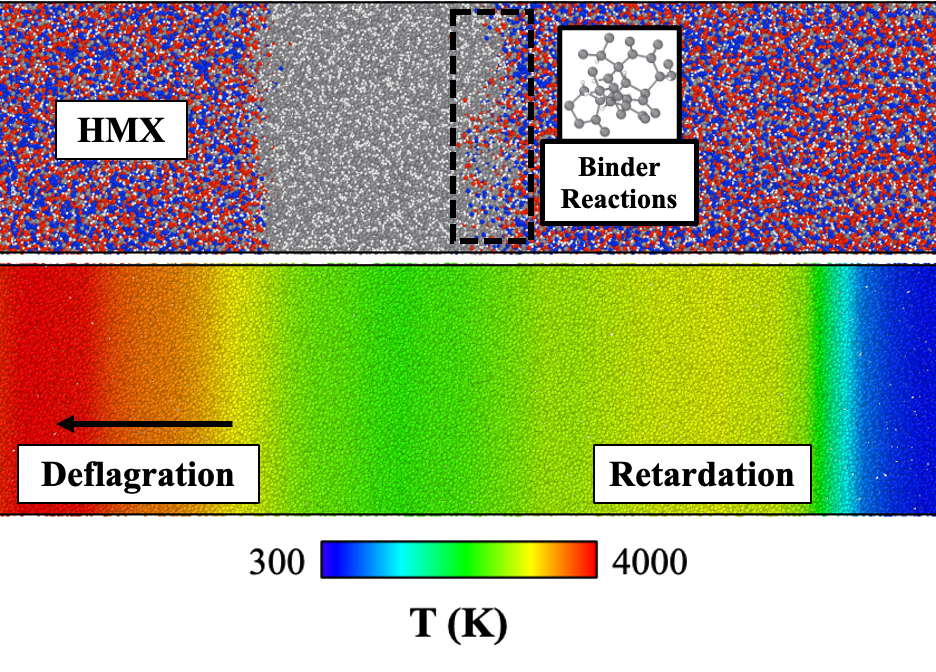}
\end{tocentry}

\end{document}